\documentstyle[aps,preprint,epsfig]{revtex}
\tighten

\begin{document}
\preprint{ TU-765 } 
\title{Gravitino Production from Heavy Moduli Decay
and Cosmological Moduli Problem Revived}

\author{Shuntaro Nakamura and Masahiro Yamaguchi} 

\address{Department of Physics, Tohoku University, Sendai,
980-8578, Japan} 

\maketitle

\begin{abstract}%
  The cosmological moduli problem for relatively heavy moduli fields
  is reinvestigated. For this purpose we examine the decay of a
  modulus field at a quantitative level. The modulus dominantly decays
  into gauge bosons and gauginos, provided that the couplings among
  them are not suppressed in the gauge kinetic function. Remarkably
  the modulus decay into a gravitino pair is unsuppressed generically,
  with a typical branching ratio of order 0.01. Such a large gravitino
  yield after the modulus decay causes cosmological difficulties. The
  constraint from the big-bang nucleosynthesis pushes up the gravitino
  mass above $10^5$ GeV. Furthermore to avoid the over-abundance of
  the stable neutralino lightest superparticles (LSPs), the gravitino
  must weigh more than about $10^6$ GeV for the wino-like LSP, and
  even more for other neutralino LSPs. This poses a stringent
  constraint on model building of low-energy supersymmetry.
\end{abstract}

\newpage

Moduli stabilization is one of the long standing problems in the
efforts to connect superstring theory to the real world. Recent
development of the flux compactifications \cite{flux} implies that some of the
moduli as well as the dilaton are stabilized at ultra high energy
scale close to the Planck scale.  However others will remain light
compared to the Planck mass. Their masses are expected to be not far
from the electroweak scale in the context of low-energy supersymmetry.
The remaining moduli will play important roles in phenomenology and
cosmology.

In fact, as is termed the cosmological moduli problem~\cite{m-problem}, such
remaining moduli would be cosmological embarrassment. It is likely
that the moduli fields have Planck scale amplitudes in the early
universe and thus their coherent oscillation would dominate the energy
density of the universe.  Since their interactions are very weak,
typically suppressed by the Planck mass, they are long-lived. If the
masses are around the electroweak scale they decay much later than 1
sec, releasing huge entropy. This will completely upset the success of
the primordial nucleosynthesis, one of the most important triumphs of
the big-bang cosmology.

A resolution of the cosmological moduli problem is to invoke a
relatively heavy moduli with mass of $10^5$ GeV or
more~\cite{MYY,Randall-Thomas,Kawasaki:1995cy,Moroi-Randall}.\footnote{
  Other ways out have also been proposed. The moduli fields may be
  fixed at a enhanced symmetry point so that the initial amplitude of
  the moduli oscillation may be small~\cite{Dine:1995uk}. Late time
  entropy production\cite{Lazarides:1985bj,Lyth:1995ka} may take place
  to dilute the unwanted relics.  It has been suggested recently that
  the moduli fields may decay rapidly in the thermal
  bath~\cite{Yokoyama:2006wt}. Finally one may seek for string
  compactifications with all moduli being stabilized at a scale close
  to the string scale.} 
Then the moduli fields decay before the
nucleosynthesis commences and will not spoil it. It has been recently
recognized that such heavy moduli masses can naturally be realized
when the moduli have superpotential of exponential type, which are
typically generated by some non-perturbative effects. For instance the
superpotential of one exponential with a constant (the KKLT-type
superpotential~\cite{KKLT}) gives the modulus mass about $8\pi^2$
heavier than the gravitino
mass~\cite{Choi:2004sx,Choi:2005ge,Endo:2005uy,Choi:2005uz,Falkowski:2005ck},
whereas the superpotential of two exponentials (race-track
superpotential \cite{racetrack}) gives an additional factor of $8
\pi^2$ to the modulus mass~\cite{BHLR}.

Thus a natural setting will be that the moduli have
masses by several orders of magnitude larger than the gravitino. The
gravitino can again be heavier than the superparticles in the minimal
supersymmetric standard model (MSSM), depending on how supersymmetry
breakdown is mediated to the MSSM sector.

The purpose of this paper is to reinvestigate cosmological issues of
the heavy moduli fields. To this end, we will examine the decays of
a modulus field, in particular, into a gravitino pair at a
quantitative level. A special attention is paid to the helicity $\pm
1/2$ components of the gravitino. We will show that the decay
amplitude is proportional to the F-auxiliary field expectation value
of the modulus field. For the heavy modulus field this is suppressed
by the ratio of the gravitino mass to the modulus mass. However, the
net result does not suffer from any suppression factor in general. A
typical branching ratio of the modulus decay into gravitinos is at a
\% level, which is natural from the counting of degrees of freedom. A
previous estimate used in 
Refs.~\cite{Hashimoto:1998mu,Kohri:2004qu,Kohri:2005ru}
is not applicable for a general case.

The non-suppressed branching ratio of the decay into the gravitinos
has striking impacts to cosmology. First of all, given a large
gravitino yield the gravitino decay would spoil the success of the
big-bang nucleosynthesis. This reincarnation of the gravitino
problem~\cite{g-problem} implies that the gravitino should weigh more
than $10^5$ GeV. A severer constraint on the gravitino mass will be
obtained, however, by the over-abundance of the neutralino dark
matter, provided that one of the neutralinos in the minimal
supersymmetric standard model (MSSM) is the lightest superparticle of
the whole theory and the R-parity is conserved so that the LSP is
absolutely stable. The requirement that the neutralino LSP abundance
does not exceed the observation leads the lower bound of the gravitino
mass of the order $10^6$ GeV for a weak scale LSP mass when the LSP is
wino-like so that the annihilation is most effective. On the other
hand, the neutralinos produced by the gravitino decay will get
thermalized if the gravitino mass is larger than $10^7$ GeV.

Such a heavy gravitino is very embarrassing when one attempts to
realize weak scale superparticle masses in the MSSM sector. We will
speculate possible resolutions to this problem at the end of the
paper.

The rest of the paper will be devoted to detail  the aforementioned
results.

\

We begin by reviewing some properties of moduli fields. In the following, we assume, for simplicity, that there is one modulus field under consideration.
As was
mentioned previously, it is a scalar field in 
a chiral supermultiplet. We denote it  by $X$. Its
properties are governed by the K\"ahler potential $K(X, X^*)$, a real
function of $X$ and its complex conjugate field $X^*$, and the
superpotential $W(X)$, a holomorphic function of $X$. In the following
we will use the Planck unit where the reduced Planck scale $M_{\rm
Pl}\simeq 2.4 \times 10^{18}$ GeV is set to unity unless otherwise
stated. The kinetic term of the $X$ field is given by
\begin{eqnarray}
  K_{X X^*} \partial_{\mu} X^* \partial^{\mu} X
\label{eq:X-kinetic-term}
\end{eqnarray}
where the indices $X$ and $X^*$ in the K\"ahler metric $K_{X X^*}$
represent the partial derivatives of $K$ with respect to $X$ and $X^*$,
respectively. In string models, one often obtains the K\"ahler potential 
of the type
\begin{eqnarray}
    K(X, X^*)=-n \log(X+X^*)
\label{eq:K-string}
\end{eqnarray}
where $n$ is an (integer) constant. In this case 
the K\"aher metric becomes
\begin{eqnarray}
   K_{X X^*}=\frac{n}{(X+X^*)^2}.
\end{eqnarray}
  The
$F$-auxiliary field of the $X$ is given by
\begin{eqnarray}
     F_X= -e^{K/2}(K_{X X^*})^{-1}(W_X+K_X W)^*
\label{eq:F-X}
\end{eqnarray}
When it is much heavier than the gravitino mass 
$m_{3/2}=\langle e^{K/2}W \rangle$, the modulus mass is given by
\begin{eqnarray}
   m_X=\langle e^{K/2}(K_{X X^*})^{-1} W_{XX} \rangle 
= -\left\langle \frac{\partial F_X^{\, *}}{\partial X} \right \rangle,
\label{eq:m-X}
\end{eqnarray}
where $\langle \cdots \rangle$ stands for a vacuum expectation value
(VEV). Supergravity corrections of order $m_{3/2}$ have been
neglected. Finally, as is known well, 
 the stationary condition of the potential with
respect to the $X$ field implies that the VEV of $F_X$ is generically 
of order $m_{3/2}^2/m_X$ (in the Planck unit) when the $X$
field takes a VEV around the Planck scale.  Whether it really survives
non-zero or not will be model dependent. For interesting cases such
as the KKLT-model and the two race-track model, the VEV of $F_X$ field
is non-vanishing and indeed of the order $m_{3/2}^2/m_X$. In the
following we assume this is the case for the modulus field $X$ under
consideration.

We now discuss  various decay modes of the modulus field.
Let us first consider the  decay into gauge bosons. The coupling
of the modulus field to vector supermultiplets ({\it i.e.,} gauge
bosons and gauginos) is conducted by the gauge kinetic function
$S(X)$.  The relevant terms in the
Lagrangian are written
\begin{eqnarray}
    {\mathcal L }_{Xgg} = -\frac{1}{4} S_R(X) F^a_{\mu \nu} F^{a \mu \nu} 
                        - \frac{1}{8} S_I (X) \epsilon^{\mu \nu \rho \sigma} F^a_{\mu \nu} F^a_{\rho \sigma}, 
\label{eq:L-gauge-bosons}
\end{eqnarray}
where $S_R$ and $S_I$ are real and imaginary parts of the gauge
kinetic function, respectively. In the above, we have taken the $S(X)$
to be universal for all gauge groups, for simplicity. Expanding
Eq.~(\ref{eq:L-gauge-bosons}) around the VEV of the $X$ field $\langle
X \rangle$, we find
\begin{eqnarray}
  {\mathcal L}_{Xgg} & =& 
 - \frac{1}{4} \langle S_R \rangle F^a_{\mu \nu} F^{a \mu \nu} 
 - \frac{1}{4} \left \langle \left( \frac{\partial S}{\partial X} \right)_R 
                                \right \rangle 
                        \delta X_R F^a_{\mu \nu} F^{a \mu \nu} 
 + \frac{1}{4} \left \langle \left(
            \frac{\partial S}{\partial X} \right)_I \right \rangle 
                        \delta X_I F^a_{\mu \nu} F^{a \mu \nu} 
\nonumber \\ 
&   & - \frac{1}{8} \langle S_I \rangle 
       \epsilon^{\mu \nu \rho \sigma} F^a_{\mu \nu} F^a_{\rho \sigma}
  - \frac{1}{8} \left\langle \left(
         \frac{\partial S}{\partial X} \right)_I \right\rangle
              \delta X_R \epsilon^{\mu \nu \rho \sigma} 
             F^a_{\mu \nu} F^a_{\rho \sigma}
   - \frac{1}{8} \left\langle \left( 
            \frac{\partial S}{\partial X} \right)_R \right\rangle
   \delta X_I \epsilon^{\mu \nu \rho \sigma} F^a_{\mu \nu} F^a_{\rho \sigma},  
\end{eqnarray}
where $\delta X \equiv X-\langle X \rangle$. 
%Recalling the kinetic
%term of the $X$ field (\ref{eq:X-kinetic-term}), we note that two real
%scalar fields which are canonically normalized, $\phi_R$ and $\phi_I$
%should be
%\begin{eqnarray}
%       \phi_R \equiv (2K_{X X^*})^{1/2} \delta X_{R}, ~~~ 
%        \phi_I \equiv (2K_{X X^*})^{1/2} \delta X_{I}.
%\end{eqnarray}
It is straightforward to compute the decay width to the gauge boson pairs. The result is
\begin{eqnarray}
      \Gamma(X_R \to g g  ) = \Gamma(X_I \to g g) 
                                     = \frac{N_G}{128 \pi} d_g^{\, 2} 
                          \frac{m^3_X}{M_{\rm Pl}^2},
\label{eq:decay-width-gauge-bosons}
\end{eqnarray}
where $d_g$ is a dimensionless constant of order unity defined (in the
Planck unit) by
\begin{eqnarray}
   d_g 
\equiv \, \, \langle K_{X X^*} \rangle^{-\frac{1}{2}} \langle S_R \rangle^{-1} 
               \bigg| \bigg\langle \frac{\partial S}{\partial X} 
                      \bigg\rangle \bigg|, 
\end{eqnarray}
and $N_G$ is the number of the gauge bosons. $N_G=12$ for
the minimal supersymmetric standard model. In deriving the above result, we 
have rescaled the field $\delta X$ and the gauge fields $A_{\mu}$ into canonically 
normalized ones.  In
Eq.~(\ref{eq:decay-width-gauge-bosons}), the reduced Planck scale has
been explicitly written. To give an example of  $d_g$, let us consider the
K\"ahler potential of the form (\ref{eq:K-string}) and the gauge kinetic
function $S(X)=X$. Then $d_g=2/\sqrt{n}$, and in fact it is of order
unity.

Evaluation of the decay width into the gaugino pairs can be done
similarly. We denote the gaugino fields by $\lambda^{(a)}$ and 
$\bar{\lambda}^{(a)}$ in two component formalism.  The relevant terms of the 
Lagrangian in this case are
\begin{eqnarray}
  {\mathcal L}_{X \tilde g \tilde g}
&=& \, \, 
 \frac{i}{2} S_R(X) \bigg[
  \lambda^{(a)} \sigma^{\mu} \tilde{\mathcal{D}}_{\mu}
  \bar{\lambda}^{(a)} + \bar{\lambda}^{(a)} \bar{\sigma}^{\mu}
  \tilde{\mathcal{D}}_{\mu} \lambda^{(a)} \bigg] - \frac{1}{2} S_I (X)
  \tilde{\mathcal{D}}_{\mu} \left[ \lambda^{(a)} \sigma^{\mu}
    \bar{\lambda}^{(a)} \right] \nonumber 
\\ 
& & + \frac{1}{4}
  \frac{\partial S}{\partial X} F_X \lambda^{(a)} \lambda^{(a)} +
  \frac{1}{4} \left( \frac{\partial S}{\partial X} F_X \right)^{*}
  \bar{\lambda}^{(a)} \bar{\lambda}^{(a)},
\label{eq:L-gauginos}
\end{eqnarray}
where the covariant derivative is defined as
\begin{eqnarray} 
\tilde{\mathcal{D}}_{\mu} \lambda^{(a)} 
  = \, \, \partial_{\mu} \lambda^{(a)}                                     
      + \frac{1}{4} \left( K_j \partial_{\mu} \phi^j - 
          K_{j^{\ast}} \partial_{\mu} \phi^{\ast j} \right) \lambda^{(a)} 
       +\cdots.
\end{eqnarray}
Utilizing the equations of motion for the gauginos, one finds that the
first line of Eq.~(\ref{eq:L-gauginos}) makes small contributions to the
decay amplitude, which are suppressed by the small gaugino masses.
On the other hand, the contributions from the second line are unsuppressed.
In fact
\begin{eqnarray}
    \frac{\partial S}{\partial X} F_X 
&= & \left\langle \frac{\partial S}{\partial X} F_X \right\rangle
  + \left\langle \frac{\partial}{\partial X} 
        \left( \frac{\partial S}{\partial X} F_X \right) \right\rangle
      \delta X
   + \left\langle \frac{\partial}{\partial X^*} 
        \left( \frac{\partial S}{\partial X} F_X \right) \right\rangle
      \delta X^*
\nonumber \\
& = & \left\langle \frac{\partial S}{\partial X} F_X \right\rangle
  + \left\langle 
       \frac{\partial^2 S}{\partial X^2} F_X +
      \frac{\partial S}{\partial X} \frac{\partial F_X}{\partial X} 
        \right\rangle
     \delta X
   + \left\langle          
   \frac{\partial S}{\partial X} \frac{\partial F_X}{\partial X^*} 
       \right\rangle
      \delta X^*
\nonumber \\
& \simeq & \left\langle \frac{\partial S}{\partial X} F_X \right\rangle
  - \left\langle 
      \frac{\partial S}{\partial X}  \right\rangle  m_{3/2}
     \delta X
   - \left\langle 
        \frac{\partial S}{\partial X} 
       \right\rangle m_{X}
      \delta X^*,
\end{eqnarray}
where use of Eqs.~(\ref{eq:F-X}) and (\ref{eq:m-X}) is made to obtain
the last equality. In the case where $m_X \gg m_{3/2}$, the decay width is
simply given\footnote{The contributions from the $F$-auxiliary part have 
not been discussed in Ref.~\cite{Moroi-Randall}.}
\begin{eqnarray}
   \Gamma( X_R \rightarrow \tilde g \tilde g)=
   \Gamma( X_I \rightarrow \tilde g \tilde g) =
                                     \frac{N_G}{128 \pi} d_g^{\, 2} 
                          \frac{m^3_X}{M_{\rm Pl}^2}.
\label{eq:decay-width-gauginos}
\end{eqnarray}
Notice that it is identical to the decay width to the gauge bosons. When
the gravitino mass is comparable to the modulus mass, the above result is
modified, but remains the same order of magnitude.

The two-body decays of the modulus field into Standard Model fermion
pairs as well as sfermions can be shown to be suppressed by powers of
the masses of final states by using their equations of
motion~\cite{Moroi-Randall}.\footnote{ For the decays into sfermion
  pairs, there are also contributions which are suppressed by powers
  of $m_{3/2}/m_X$. These are irrelevant as the gravitino mass is
much smaller than the modulus mass.} The three-body decays such as
quark-quark-gluon will not receive this chiral suppression. However
they will be suppressed by $\alpha_s/\pi$ and they will not dominate
over the decays into the gauge bosons and gauginos.  Thus we discard
them in the subsequent discussion.  On the other hand, terms like
$X^{\dagger} H_u H_d$ in the K\"ahler potential, with $H_u$ and $H_d$
being the Higgs multiplets in the MSSM, may make sizable contribution
to the decay width, which can be comparable to the gauge and gaugino
final states. Whether these terms exist or not are quite model
dependent, and we will not consider them here.

We now examine the modulus decay into the gravitino pair. In the
Unitary gauge, the relevant interaction terms are in the gravitino
bilinear terms in the supergravity Lagrangian
\begin{eqnarray}
 {\mathcal L}_{3/2} = -\epsilon^{\mu \nu \rho \sigma} \bar{\psi}_{\mu} \bar{\sigma}_{\nu} 
\tilde{\mathcal{D}}_{\rho}\psi_{\sigma} 
 -e^{K/2}W^* \psi_{\mu}\sigma^{\mu \nu}\psi_{\nu}
-e^{K/2}W \bar{\psi}_{\mu}\bar{\sigma}^{\mu \nu}\bar{\psi}_{\nu},
\label{eq:L-gravitino}
\end{eqnarray}
where $\psi_{\mu}$ stands for the gravitino in two component formalism,
and the covariant derivative is given by
\begin{eqnarray}
    \tilde{\mathcal{D}}_{\rho}\psi_{\sigma}
   \equiv \partial_{\rho}\psi_{\sigma}
      +\frac{1}{4}(K_j
      \partial_{\rho}\phi^j-K_{j^*}\partial_{\rho}\phi^{*j})
      \psi_{\sigma} + \cdots.
\end{eqnarray}
Making  a field-dependent chiral transformation
\begin{eqnarray}
   \psi_{\mu} \mapsto \left(  \frac{W}{W^*}  \right)^{-1/4}\psi_{\mu},
\label{eq:chiral-transformation}
\end{eqnarray}
Eq.~(\ref{eq:L-gravitino}) reduces to 
\begin{eqnarray}
  -\epsilon^{\mu \nu \rho \sigma} \bar{\psi}_{\mu} \bar{\sigma}_{\nu} 
\partial_{\rho}\psi_{\sigma} 
-\frac{1}{4}\epsilon^{\mu \nu \rho \sigma} 
(G_j \partial_{\rho}\phi^j-G_{j^*}\partial_{\rho}\phi^{*j})
\bar{\psi}_{\mu}  \bar{\sigma}_{\nu}\psi_{\sigma} 
 -e^{G/2}(\psi_{\mu}\sigma^{\mu \nu}\psi_{\nu}
   +\bar{\psi}_{\mu}\bar{\sigma}^{\mu \nu}\bar{\psi}_{\nu}),
\label{eq:L-gravitino-modified}
\end{eqnarray}
where $G$ is the total K\"ahler potential defined by
\begin{eqnarray}  
    G(X,X^*)=K(X,X^*)+\log|W(X)|^2
\end{eqnarray}
In this convention, the gravitino mass is $m_{3/2}=\langle e^{G/2} \rangle$. By
expanding the Lagrangian (\ref{eq:L-gravitino-modified})
in terms of $\delta X$ and $\delta X^*$, one finds
the interaction terms
\begin{eqnarray}
 & &-\frac{1}{4}\epsilon^{\mu \nu \rho \sigma} 
(\langle G_X \rangle \partial_{\rho} \delta X 
 -\langle G_{X^*} \rangle \partial_{\rho} \delta X^{*})
\bar{\psi}_{\mu}  \bar{\sigma}_{\nu}\psi_{\sigma}
\nonumber \\
& & 
 - \frac{1}{2} m_{3/2}
  (\langle G_X \rangle \delta X + \langle G_{X^*} \rangle \delta X^{*})
  (\psi_{\mu}\sigma^{\mu \nu}\psi_{\nu}
   +\bar{\psi}_{\mu}\bar{\sigma}^{\mu \nu}\bar{\psi}_{\nu}).
\end{eqnarray}
We note that the coupling is governed by $\langle G_X \rangle $. It is
related to the auxiliary field expectation value as follows:
\begin{eqnarray}
    \langle F_{X} \rangle =- \langle (G_{X X^*})^{-1} e^{G/2}G_{X^*} \rangle
\end{eqnarray}
Since the  VEV of $F_X$ is naturally  of order $m_{3/2}^2/m_X$, one expects 
\begin{eqnarray}
       \langle G_X \rangle \sim m_{3/2}/m_X
\end{eqnarray}
in the Planck unit, which we assume to be the case in the
following discussion. We should stress that this is indeed the case for the
KKLT-type set-up and also for the race-track supersymmetry breaking
scenario.  

We are now at the position to compute the decay width into the
gravitino pair. It is evident that the helicity $\pm 1/2$ components
will give dominant contributions if they are non-vanishing, because
they will contain enhancement factor of $1/m_{3/2}$. To see the point,
let us take, for simplicity, $ \langle G_X \rangle$ to be real and
consider the decay of the real component of $\delta X$. Then the decay
amplitude for a given set of helicity components is written\footnote{
  The chiral transformation (\ref{eq:chiral-transformation}) makes the
  expression of the matrix element very simple. If we use the original
  Lagrangian (\ref{eq:L-gravitino}), then the amplitude contains more
  than one term. In this case we checked that partial cancellation
  takes place between different terms, arriving at the same result we 
  describe here.}
\begin{eqnarray}
    {\mathcal M}(X_R \to \psi_{3/2} \psi_{3/2})
   =i \frac{1}{\sqrt{2}}\langle G_{X X^*} \rangle ^{-1/2}
      \langle e^{G/2} G_X \rangle \bar{v}_{\mu} (k') u^{\mu}(k),
\end{eqnarray}
where $u_{\mu}$ and $v_{\mu}$ are gravitino wave functions (in four
component formalism). To derive this, we have used that the gravitino is
a Majorana fermion and the two wave functions are related by the
Majorana condition $v_{\mu}=C\bar{u}^T_{\mu}$ with $C$ being the charge
conjugation matrix. The wave functions of a massive spin $3/2$ field are
conveniently expressed as a tensor product of a vector and a 
spinor. For instance, 
the helicity 1/2 component is written
\begin{eqnarray}
            u_{\mu}(k;1/2)=\sqrt{\frac{2}{3}}\epsilon_{\mu}(k;0)u(k;1/2)
                         + \sqrt{\frac{1}{3}}\epsilon_{\mu}(k;1)u(k;-1/2)
\end{eqnarray}
with self-explanatory notation. Here 
$\epsilon_{\mu}(k,0)\simeq k_{\mu}/m_{3/2}$ at a high-energy limit. 
The decay amplitude into
the two helicity $1/2$ components of the gravitino is thus expressed as
\begin{eqnarray}
      {\mathcal M}_{+1/2} & = & {\mathcal M}(X_R \to \psi_{3/2}(1/2)
 \psi_{3/2}(1/2))
 \nonumber \\
& \simeq & i \frac{1}{\sqrt{2}}\langle G_{X X^*} \rangle ^{-1/2}
      \langle e^{G/2} G_X \rangle 
      \epsilon_{\mu}^* (k';0) \epsilon^{\mu}(k;0) 
      \bar{v}(k';1/2) u(k;1/2)
 \nonumber \\
& \simeq & i \frac{1}{\sqrt{2}}\langle G_{X X^*} \rangle ^{-1/2}
      \langle e^{G/2} G_X \rangle
      \frac{k \cdot k'}{m_{3/2}^2} 
      \bar{v}(k';1/2) u(k;1/2)
\nonumber \\
& \simeq & -i \frac{1}{3\sqrt{2}}d_{3/2} m_{X}^2
\end{eqnarray}
where  $d_{3/2}$ is defined as follows:
\begin{eqnarray}
         \langle G_{X X^*} \rangle^{-1/2} \langle e^{G/2} G_X \rangle
       \equiv d_{3/2}\frac{m_{3/2}^2}{m_X}.
\end{eqnarray} 
As was seen previously, $d_{3/2}$ is a dimensionless constant of order unity.
Notice that the l.h.s. of the above is the $F$-auxiliary field of the
canonical normalized supermultiplet. The same expression is obtained for 
the decay into the helicity $-1/2$ components. On the other hand, 
the decays into the helicity $\pm 3/2$
components are suppressed by powers of $m_{3/2}/m_X$. 
Thus the decay width is computed to be
\begin{eqnarray}
     \Gamma(X_R \to \psi_{3/2} \psi_{3/2}) 
        =\frac{1}{288 \pi} d_{3/2}^2 \frac{m_{X}^3}{M_{\rm Pl}^2}
        \label{eq:decay-width-to-gravitino}
\end{eqnarray}
at the limit $m_{3/2} \ll m_X$.
Here the reduced Planck scale has been written explicitly in the final
expression of the decay width. 
A computation can also be performed for the imaginary part of the $\delta
X$, with the same result as Eq.~(\ref{eq:decay-width-to-gravitino}).
%\begin{eqnarray}  
%\Gamma(X_I \to \psi_{3/2} \psi_{3/2}) =
%\frac{1}{288 \pi} d_{3/2}^2 \frac{m_{X}^3}{M_{\rm Pl}^2}.
%\end{eqnarray}

Thus the modulus
field dominantly decays into the gauge bosons and the gauginos with the
total decay width
\begin{eqnarray} 
      \Gamma_{\rm tot} & \equiv & \Gamma (X \to {\rm all})
 \simeq \Gamma (X \to g g )
                     + \Gamma (X \to \tilde g \tilde g)
\nonumber \\
 &    = &  \frac{N_G}{64 \pi} d_g^{\, 2} 
                          \frac{m^3_X}{M_{\rm Pl}^2}
= \frac{3}{16 \pi}  \left( \frac{N_G}{12} \right) d_g^{\, 2} 
                           \frac{m^3_X}{M_{\rm Pl}^2}
\end{eqnarray}
and the branching ratio of the decay to the gravitino pair
\begin{eqnarray}
    B_{3/2} \equiv  {\rm Br} (X \to \psi_{3/2} \psi_{3/2})
 =    \frac{1}{54} \left( \frac{N_G}{12} \right)^{-1} 
         \frac{d_{3/2}^{\, 2}}{d_g^{\, 2}}.
\end{eqnarray}
With $d_{g}$ and $d_{3/2}$ being the constants of order unity, we find
that the
branching ratio to the gravitinos is of order $10^{-2}$.

The production of the gravitinos at the modulus decay has striking
impacts on the cosmology. Here we consider the situation where the
coherent oscillation of the modulus field will dominate the energy
density of the universe after primordial inflation, releasing huge 
entropy and reheating the universe at the decay. The reheating
temperature at the modulus decay is estimated by equating the total
decay width to the expansion rate of the universe at the reheating:
\begin{eqnarray}
 T_R &=& \left( \frac{90}{\pi^2 g_*(T_R)} \right)^{1/4}
         \sqrt{\Gamma_{\rm tot} M_{\rm Pl}}
\nonumber \\
     &=& 4.9 \times 10^{-3} \left( \frac{g_*(T_R)}{10} \right)^{-1/4}
              \left( \frac{N_G}{12} \right)^{1/2}
              d_{g} 
             \left( \frac{m_X}{10^5 \rm{GeV}} \right)^{3/2} {\rm GeV},
\end{eqnarray}
where $g_*(T_R)$ is the effective degrees of freedom of the radiation
at the reheating. The gravitino yield produced by the modulus decay,
which is defined by the ratio of the gravitino number density
$n_{3/2}$ relative to the entropy density $s$, can easily be evaluated
as
\begin{eqnarray}
         Y_{3/2} &\equiv& \frac{n_{3/2}}{s}
\nonumber \\
          & = & \frac{3}{2} B_{3/2} \frac{T_R}{m_X}
\nonumber \\
          & = &
 0.73 \times 10^{-7} B_{3/2} \left( \frac{g_*(T_R)}{10} \right)^{-1/4}
              \left( \frac{N_G}{12} \right)^{1/2}
              d_{g} 
             \left( \frac{m_X}{10^5 \rm{GeV}} \right)^{1/2} .
\label{eq:gravitino-yield}
\end{eqnarray}       

A constraint on the gravitino yield comes from the big-bang
nucleosynthesis (BBN). As is well-known, the success of the BBN would
be threatened by the electromagnetic showers (as well as the hadronic
showers) produced at the gravitino decay. It is termed the gravitino
problem. The gravitinos are produced via scattering processes in the
thermal bath after primordial inflation epoch.  In the situation we
are considering, the gravitinos in this origin are diluted by the
entropy production at the modulus decay. However they are regenerated
directly by the modulus decay. The requirement that the gravitino
decay products should not spoil the BBN severely constrains the
gravitino abundance. Recent 
analyses~\cite{Kawasaki:2004yh,Kawasaki:2004qu,Kohri:2005wn}
show that, for the gravitino mass in the
range $m_{3/2} \simeq 10^3 - 10^4$ GeV, the constraints from D/H and
$^6$Li are the severest, leading to $Y_{3/2} \lesssim 10^{-16}$ when
the hadronic branching ratio of the gravitino decay $B_h$ is 1, and
$Y_{3/2} \lesssim 10^{-13}$ when $B_h =10^{-3}$.  Comparing these
numbers with the gravitino yield obtained at the modulus decay
(\ref{eq:gravitino-yield}), one sees that the latter exceeds by
several orders of magnitude. For lighter gravitino $m_{3/2} \simeq
10^2-10^3$ GeV, $^3$He/D also plays a role.  The constraint in this
range remains very severe, roughly speaking at the level $Y_{3/2}
\lesssim 10^{-16}$.  It becomes somewhat weaker for heavier gravitino,
especially for the case $B_h =10^{-3}$. However the yield
(\ref{eq:gravitino-yield}) still exceeds the constraint from $^4$He
abundance. Finally the constraint disappears when the mass is above
$10^5$ GeV, corresponding to the life-time shorter about $10^{-2}$
sec. Thus we conclude that the (unstable) gravitino whose mass is less
than $10^{5}$ GeV is excluded by the BBN constraint.

Though we already obtained a very severe bound on the gravitino mass,
this is not the end of the story. When the gravitino is unstable, it
decays to lighter superparticles, {\it i.e.} $R$-parity odd particles.
Under the assumption of $R$-parity conservation, the lightest
superparticle (LSP) is absolutely stable. A plausible candidate for
the LSP in the MSSM is the lightest in the neutralino sector, that is
a linear combination of the neutral gauginos and higgsinos.  The
neutralino LSP, if stable, will contribute as (a part of)  the cold dark
matter whose abundance is bounded from cosmological observations. The
second question we would like to address is therefore whether the neutralino
abundance produced by the gravitino decay will not exceed the upperbound of
the dark matter inferred by the observations. We should note here that the
 problem of the over-abundance of the neutralinos produced by the moduli
 decay was addressed in Refs.~\cite{MYY,Kawasaki:1995cy}.

The gravitinos produced at the modulus decay do not interact with
others. Thus the yield of the gravitinos does not change until they
decay. The gravitino decay width into the MSSM particles is 
(see for instance Ref.~\cite{Moroi:1995fs})
\begin{eqnarray}
    \Gamma_{3/2}=\frac{193}{384 \pi} \frac{m_{3/2}^{3}}{M_{\rm Pl}^2},
\label{eq:gravitino-width}
\end{eqnarray}
when all MSSM (super)particles are included in the final states
and their masses are neglected, which is justified for the gravitino
 much heavier than the MSSM (super)particles. It follows from 
Eq.~(\ref{eq:gravitino-width}) that the
temperature of the universe at the gravitino decay is
\begin{eqnarray}
    T_{3/2} \simeq \left( \frac{90}{\pi^2 g_*(T_{3/2})} \right)^{1/4}
           \sqrt{\Gamma_{3/2} M_{\rm Pl}}.
\end{eqnarray}
Numerically it reads
\begin{eqnarray}
   T_{3/2} \simeq 7.9 \times 10^{-3} 
       \left( \frac{g_* (T_{3/2})}{10} \right)^{-1/4}
       \left( \frac{m_{3/2}}{10^5 {\rm GeV}} \right)^{3/2} {\rm GeV}
\end{eqnarray}

The gravitino decays into lighter superparticles, followed by
cascade decays to the neutralino LSP. The neutralino LSPs produced this
way are so abundant  that they annihilate with each other. 
The annihilation process terminates when the annihilation rate reduces to
the expansion rate of the universe at the gravitino decay $H(T_{3/2})$~\cite{MYY}
\begin{eqnarray}
         \langle \sigma_{\rm ann} v_{\rm rel} \rangle n_{\chi}
       \simeq H(T_{3/2}),
\end{eqnarray}
where $\sigma_{\rm ann}$ is the annihilation cross section of the two
neutralino LSPs, $v_{\rm rel}$ their relative velocity, $\langle
\cdots \rangle$ represents the average over the LSP momentum
distribution, and $n_{\chi}$ is the number density of the neutralino LSPs. The above 
argument derives  an estimate for the neutralino abundance:
\begin{eqnarray}
  \left. \frac{n_{\chi}}{s} \right|_{T_{3/2}}
\simeq 
    \left. \frac{H(T_{3/2})}{\langle \sigma_{\rm ann} v_{\rm rel} \rangle s}
        \right|_{T_{3/2}}
  =\frac{1}{4} \left( \frac{90}{\pi^2 g_* (T_{3/2})} \right)^{1/2}
      \frac{1}{\langle \sigma_{\rm ann} v_{\rm rel} \rangle T_{3/2} M_{\rm Pl}}
\label{eq:neutralino-yield}
\end{eqnarray}
The yield remains constant until today. A more sophisticated
evaluation requires to solve Boltzmann equations numerically or
analytically~\cite{Nagano:1998aa}, but the estimate given above is
sufficient for the purpose of the present paper.

The annihilation cross section depends on the LSP component as well as
the superparticle mass spectrum. To maximize the annihilation effects
in the MSSM, let us consider the wino LSP, the neutral component of
the $SU(2)_L$ gauginos. Assuming that it is heavier than the $W$-boson,    
the dominant mode of the wino annihilation is
into $W$-boson pair via charged wino exchange.  The annihilation cross section
is computed to be \cite{Moroi-Randall}
\begin{eqnarray}
       \langle \sigma_{\rm ann} v_{\rm rel} \rangle
     = \frac{g_2^4}{2 \pi} \frac{1}{m_{\chi}^2}
       \frac{(1-x_W)^{3/2}}{(2-x_W)^2},
\label{eq:cross-section}
\end{eqnarray}
where $g_2$ is the $SU(2)_L$ gauge coupling constant, $m_{\chi}$ the
wino mass, and $x_W=m_W^2/m_{\chi}^2$.  Here possible co-annihilation
effect has  not been taken into account~\cite{Mizuta:1992qp}. 

With (\ref{eq:neutralino-yield}) and (\ref{eq:cross-section}), it is
straightforward to compute the wino relic abundance today. The result
is conveniently expressed by the relic mass density relative to the
entropy density:
\begin{eqnarray}
   \frac{m_{\chi} n_{\chi}}{s} \simeq 0.43 \times 10^{-9} {\rm GeV} 
          \frac{(2-x_W)^2}{(1-x_W)^{3/2}} 
          \left( \frac{g_*(T_{3/2})}{10} \right)^{-1/4} 
          \left( \frac{m_{\chi}}{100 {\rm GeV}} \right)^3 
          \left( \frac{m_{3/2}}{10^5 {\rm GeV}} \right)^{-3/2} ,
\end{eqnarray}
or in terms of the density parameter $\Omega_{\chi}$ which is defined by
the ratio of the LSP mass density to the critical mass density of the universe
\begin{eqnarray}
   \Omega_{\chi} h^2 \simeq 0.12 \times \frac{(2-x_W)^2}{(1-x_W)^{3/2}} 
          \left( \frac{g_*(T_{3/2})}{10} \right)^{-1/4} 
          \left( \frac{m_{\chi}}{100 {\rm GeV}} \right)^3 
          \left( \frac{m_{3/2}}{10^5 {\rm GeV}} \right)^{-3/2} ,
\end{eqnarray}
with  $h\simeq 0.72$ being the Hubble constant in units of 100 km/s/Mpc. 

\begin{figure}[htbp]
\begin{center}
\includegraphics[scale=0.7]{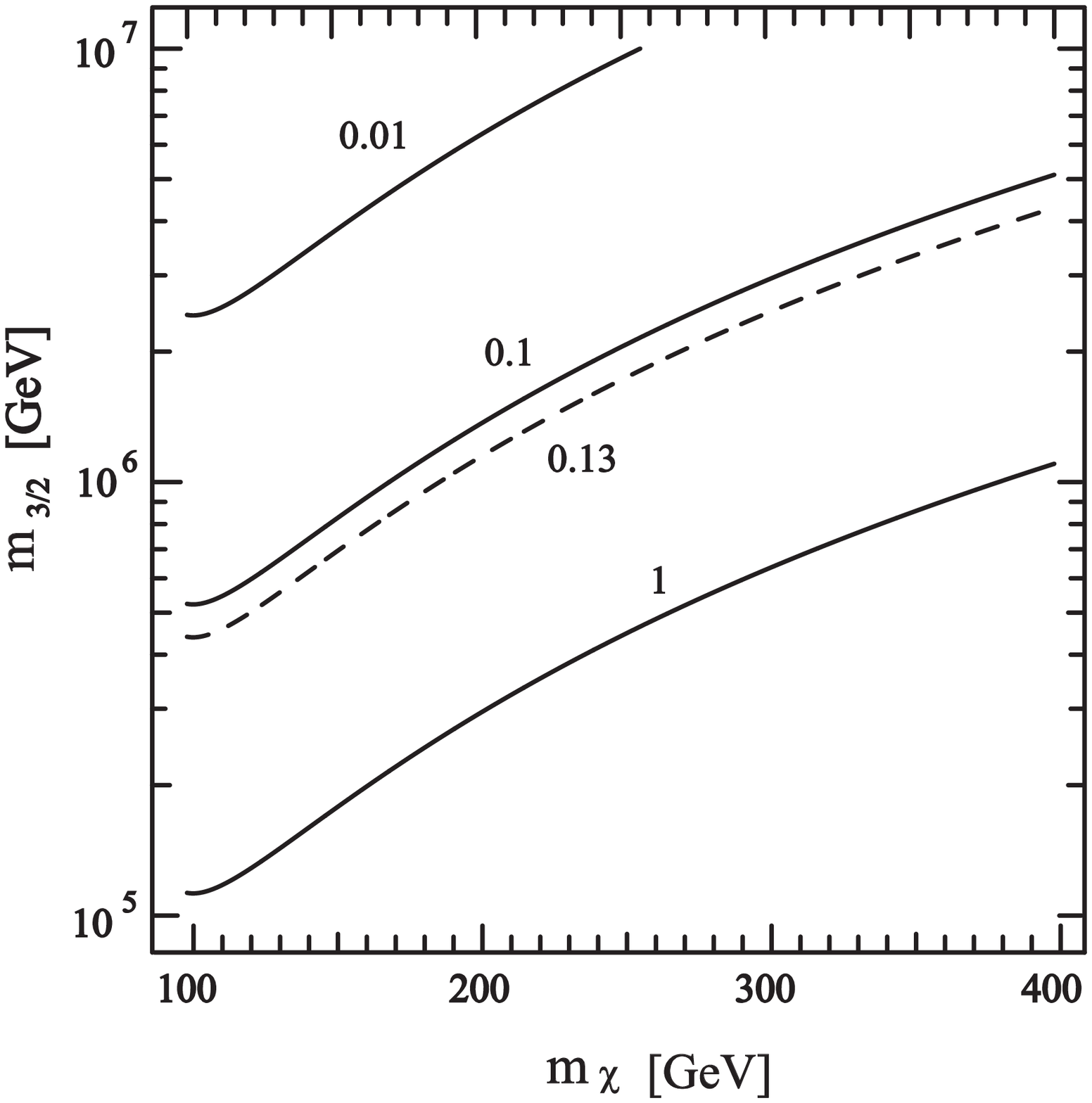}
\end{center}
\caption{ Constant contours of the density parameter
$\Omega_{\chi} h^2$ are drawn in the $m_{\chi}$-$m_{3/2}$ plane. Three
real lines represent $\Omega_{\chi} h^2=$0.01, 0.1, and 1, from the above. 
Also shown by a dashed line is  the contour  of
$\Omega_{\chi} h^2 = 0.13$, approximately corresponding to the 95\%CL
upperbound of the cold dark matter abundance from the cosmological
observations. }
\label{NK_contour}
\end{figure}

In Fig.~\ref{NK_contour}, the constant contours of the density parameter
$\Omega_{\chi} h^2$ are drawn in the $m_{\chi}$-$m_{3/2}$ plane. The
real lines represent $\Omega_{\chi} h^2=$0.01, 0.1, and 1. Given a LSP
mass, the density parameter decreases as the gravitino mass increases.
In the same figure, we also show the contour  of
$\Omega_{\chi} h^2 = 0.13$, roughly corresponding to the 95\%CL
upperbound of the cold dark matter abundance from the cosmological
observations~\cite{WMAP}. Thus in order to avoid too much abundance of
the neutralino LSPs, the region below this line should be
excluded. We find that for the wino mass in the weak scale the
gravitino should be heavier than ${\cal O}(10^6)$ GeV. On the other
hand, around the line of $\Omega_{\chi}h^2=0.1$, the wino-like LSPs
produced by the gravitino can constitute the dark matter of the
universe. Here we should caution that the estimate of the relic
abundance given in this paper is rather rough, which may contain an
error of factor 2 or so.

As we mentioned earlier, the wino case gives the largest annihilation cross
section and thus the weakest constraint on the gravitino mass.  A similar
but slightly severer bound will be obtained for the higgsino LSP case.
On the other hand, the bino LSPs will remain too much with such a low
$T_{3/2}$~\cite{MYY,Kawasaki:1995cy}. We note that when the gravitino mass becomes heavier than
about $10^7$ GeV, the produced LSPs will get thermalized and the
conventional computation of the relic abundance can apply.

Here we would like to briefly discuss what would happen if the decay
into the gravitinos is negligibly small. This is the case when the VEV
of auxiliary field $F_X$ of $X$ is (accidentally) small.  In this case
the superparticles are produced at the modulus decay with the
branching ratio of 0.5. A similar argument given for the gravitino
decay can apply except that the gravitino decay temperature should be
replaced with the reheating temperature of the modulus field. Then we
obtain quite a similar lower bound on the modulus mass to avoid the
over-abundance of the neutralino LSPs
\cite{MYY,Kawasaki:1995cy}. However there are some substantial
differences in the two cases. The modulus mass of the order $10^6$ or
$10^7$ GeV may be acceptable from model building point of view when
one considers a KKLT-type model or a racetrack model.  There are
models in which the modulus mass is separated from the supersymmetry
breaking scale. A simple model to realize this situation was given in
Ref.~\cite{Kallosh-Linde}. On the other hand, the gravitino mass is
directly related to the supersymmetry breaking scale. Thus one has to
elaborate model building to accord with a very heavy gravitino. We will
come back to this point shortly.

Let us summarize what we have obtained in this paper. We have considered the
moduli decays into various two body final states and discussed the
impacts on the cosmology.  It turns out that the total decay width is
of order $m_X^3/16 \pi M_{\rm Pl}^2$ as far as the coupling of the
modulus field to the gauge multiplets is not suppressed in the gauge kinetic function. The
main decay modes in this case are the decays into the gauge boson
pairs as well as those into the gauginos. The relevant coupling to the
latter decay modes emerge when the auxiliary component of the modulus
field is integrated.  The most important is the decay into the
gravitino pair. We have shown that the coupling of the modulus field
to the helicity $\pm 1/2$ components of the gravitino is proportional
to the VEV of the modulus auxiliary field, {\it i.e.} the
supersymmetry breaking of the modulus field.  It is known that this
VEV is of the order $m_{3/2}^2/m_{X}$ in the Planck unit for the field
with a Planck scale expectation value. Then the decay width into the
gravitino pair follows  a naive dimensional counting and its
branching ratio is at (a few) \% level. This large branching ratio has
striking impacts on the cosmology associated with the
gravitino. Firstly the daughter particles produced by the gravitino
decay would upset the BBN. What is worse, the neutralino LSPs produced
by the gravitino decays would exceed the abundances of the cold dark
matter inferred by the cosmological observations. These constraints
require the mass of the gravitino heavier than at least about $10^6$
GeV for the wino-like LSP case. For other types of neutralino LSPs,
the gravitino mass bound may reach $10^7$ GeV.

Such a heavy gravitino will cause serious difficulty to model building
of low-energy supersymmetry. In particular, superconformal anomaly
mediation~\cite{AM} will make contributions to soft masses. They are
typically proportional to the VEV of the auxiliary component of the
superconformal compensator $F_{\phi}$, suppressed by one-loop factor
of $1/16 \pi^2$.  Naively the value of $F_{\phi}$ is comparable to the
gravitino mass. If this is the case, with the gravitino mass of $10^6$
GeV or higher, the resulting soft masses would be far above the
electroweak scale, diminishing the very motivation of low-energy
supersymmetry. One may speculate that compactification with warped
metric may help, where the MSSM sector is localized on a brane with a
warp factor~\cite{Luty:2002ff} (see also Ref.~\cite{Goh:2005be}). A
modest warp factor of $10^{-1}-10^{-2}$ will be enough.  Then the VEV
of $F_{\phi}$ at the location of the MSSM brane will receive the
suppression from the warp factor compared to the gravitino mass.

\acknowledgments

The authors would like to thank T.~Asaka, T.~Moroi, Y.~Shimizu, and A.~Yotsuyanagi for 
useful discussions. The work was partially supported by the grants-in-aid  
from the Ministry
of Education, Science, Sports, and Culture of Japan, No.~16081202 and No.~17340062.

%%%%%%%%%%%%%%%%%%%%%%%%%%%%%%%%%%%%%%%%%%%%%%%%%%%%%%%%%%%%%%%%%%%%

\end{document}